\documentclass[fleqn,twoside]{article}
\usepackage{espcrc2,epsfig,dcolumn}

\newcommand{\AmS}{{\protect\the\textfont2
  A\kern-.1667em\lower.5ex\hbox{M}\kern-.125emS}}

\title{Constrained fitting of three-point functions}

\author{
  M. G\"urtler\address[a]{Department of Physics and Astronomy, University of
    Glasgow, Glasgow G12 8QQ, UK}, 
  C.T.H. Davies\addressmark[a],
  J. Hein\address[b]{Department of Physics and Astronomy, University of Edinburgh,
    Edinburgh, EH9 3JZ, UK},
  J. Shigemitsu\address[c]{Department of Physics, The Ohio State University,
    OH 43210, USA}
  and
  G. P. Lepage\address[d]{Newman Laboratory of Nuclear Science, Cornell University,
    Ithaca NY 14853, USA}
  }       
\begin{document}
\renewcommand{\topfraction}{0.5}
\renewcommand{\thefootnote}{\fnsymbol{footnote}}
\begin{abstract}
\vspace{+1mm}
  We determine matrix elements for $B \to D$ semileptonic decay. The use
  of the constrained fitting method and multiple smearings for both two- and
  three-point correlators allows an improved calculation of the form factors.
\vspace{+1mm}
\end{abstract}
\maketitle

\section{Constrained curve fitting}

The aim of this contribution is to illustrate the method of constrained curve
fitting by application to three-point functions describing the semileptonic
decay of heavy-light mesons. For a discussion of the method see \cite{lepage}.
The constraint method offers the chance to include knowledge about the fit
parameters and generally results in very stable fits. This allows for much
more ambitious fits than previous methods and the results are less prone to
systematic errors from fitting.

We apply the method to the subset ($\beta=5.7$, $a\, m_Q=2,4,8$) of the data
analysed in \cite{Hein:2000qu}, where NRQCD is used for the heavy quark.  The
lattice size is $12^3 \times 24$. The smearings used are a Gaussian with a
radius of $a\,r_Q=2.0$ for the heavy and local for the light quark (referred
to as smearing~1) and $a\,r_{q/Q}=3.0$ for both (smearing~`2). All combinations
of these smearings at source and sink were used. It has been demonstrated that
these are not the optimal smearings \cite{davies}, in particular for
nonvanishing momenta, which leaves room for improvement of the results
presented here.

\section{Form factors}
We aim at extracting the form factors $h^+(\omega)$ from the matrix elements

\begin{eqnarray*}
\left<B\left|V_0\right|D\right>&=&F_1(q^2)\; \left(p_B+p_D\right)_0\\
&=&\sqrt{m_B\,m_D}\;h^+(\omega)\;(v_B+v_D)_0
\end{eqnarray*}

for the elastic case $p_B^2=p_D^2$ with degenerate masses $m_B=m_D$.  $V_0$ is
the leading order heavy-heavy current for NRQCD~\cite{Hein:2000se}.

First we extract the form factors for each value of $q^2$.  We fit the
two-point functions parametrised by

$$
G_2^{ij}=
\sum_{k=1}^n 
A^i_k A^j_k \exp (-E_k(t-t_0))
$$ 

simultaneously with the three-point functions using the ansatz

$$G_3^{ij}=
\sum_{k,l=1}^n 
A^i_k\,J_{kl}\,A^j_l e^{-E_k\,(t-t_0)} e^{-E_l\,(T-t)}
$$

with current matrix elements $J_{kl}$, where $i,j$ label the smearings and
$k,l$ the energy levels. We set $J_{kl}=J_{lk}$, and $J_{11}$ is
$h^+(\omega)$. The distance between quark source
and sink is $12$. Instead of the energies we in fact used as parameters the logs
of energy differences as in \cite{lepage}. Our standard priors are $0.7(*/2)$ for the ground state
energy\footnote{{\em i.e.} our prior favours $\ln 0.7 \pm \ln 2$ for the fit
  parameter $\ln E_{\mathrm{g.s.}}$ } and $0.6(*/2)$ for the energy differences between
the excitations. The first is known from previous fits; the excitation energy
priors are appropriate to typical radial excitations at this lattice spacing.
For the amplitudes we took as priors $1.0(\pm1.5)$ and $0.5(\pm 1.5)$ for the
ground states, $-0.4(\pm 2)$ for $A^1_2$ and $0.1(\pm 2)$ for the rest. These
are based on experience from previous fits but allow for plenty of variation
(the typical errors on the amplitudes from the fits are $0.01 \ldots 0.05$).
The priors for the $J_{kl}$ are $1(\pm 0.5)$ for $J_{kk}$ and ${0.1(\pm 0.5)}$
for $J_{kl}$ $(k \ne l)$.

\begin{figure}[tbh]
\epsfig{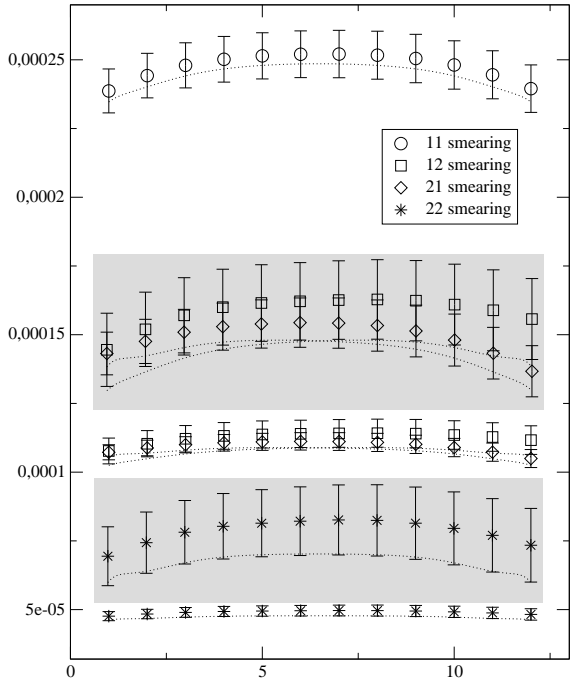}
\caption{Three-point functions and 6-exp fit results
  ($p^2=1,q^2=2,m_Q=2$); the gray boxes are (displaced)
  enlargements of the functions underneath }
\label{fig:3pt}
\end{figure}

The minimisation was done using the Levenberg-Marquardt algorithm.  The value
of $\chi^2_{\mathrm{aug}}$ for the reliable fits (i.e. using a sufficient
number of exponentials) is generally about $1$. The contribution of the
two-point functions is very small (as the examples in \cite{lepage}), the
priors contribute only marginally. Quoted errors are extracted from the
curvature matrix.  Figure~\ref{fig:3pt} shows a typical fit result for the
three-point functions.  To shed light on the effect of the priors we repeated
the calculation with priors away from the expectations and allowing for
smaller deviations.
We find that the data strongly constrain the lowest lying current matrix
elements $J_{11}$ and $J_{12}$, but have little information about the higher
ones. This is expected because off-diagonal matrix elements should be, and
are, small, because they represent the overlap (as $\bar q ^2 \to 0$) of
orthogonal states.

\begin{figure}[thb]
\epsfig{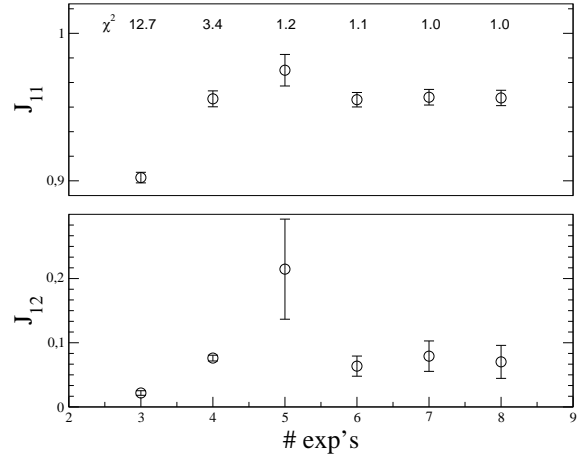}
\vspace{-1cm}
\caption{The current matrix elements for the lowest lying states vs. the number of exponentials used in the fit ($p^2=1,q^2=2,m_Q=2$)}
\label{fig:currents_vs_exps}
\end{figure}

The dependence of the lowest currents on the number of exponentials in the fit
is demonstrated in Fig.~\ref{fig:currents_vs_exps}. Stability sets in for six
exponentials.

\section{Kinetic masses}
\label{sec:kinetic-masses}

To convert these form factors into $h^+(\omega)$ one needs the kinetic masses.
They are calculated from the energy splittings $\Delta E$ between the $p^2=1$
and $p^2=0$ states $$m=(p^2-(\Delta E)^2)/(2 \Delta E).$$

Our method permits us to extract them directly from simultaneous fits of the
respective smeared-smeared two-point functions. This is an example where the
usage of the classical fitting method could be misleading if one considers the
two-point functions separately, because the two-point functions seem to
plateau much earlier than their ratio.

The values for the kinetic masses for the three heavy quark masses
investigated are listed in the following table:

\vspace{5mm}
\begin{tabular}{@{}lll}
  \hline
  $m_Q$ & $\Delta E$ & $ m_{\mathrm kin} $  \\
  \hline
  2 & 0.058(1) & 2.9(2)\\      
  4 & 0.029(3) & 4.7(5)\\      
  8 & 0.019(3) & 7.3 (1.3)\\
  \hline
\end{tabular}\\[2pt]

\begin{figure}[thb]
\epsfig{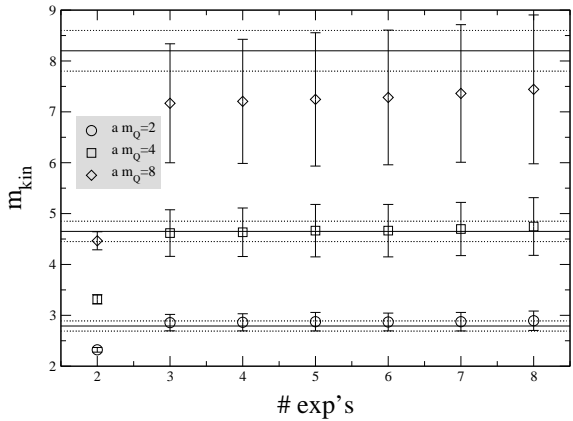}
\vspace{-8mm}
\caption{Kinetic mass as function of the number of exponentials in the fit;
  the horizontal lines are the perturbative results from \cite{Hein:2000qu}}
\label{fig:mkin_vs_exps}
\end{figure}

This much simpler fit shows stable results already using three exponentials
(see Fig.~\ref{fig:mkin_vs_exps}).  

In the future we plan to fit simultaneously all the two- and three-point
functions at different momenta and the kinetic mass will then come directly
from the same fits that give the current matrix elements.

\section{Isgur-Wise function}
\label{sec:isgur-wise-function}

\begin{figure}[thb]
\epsfig{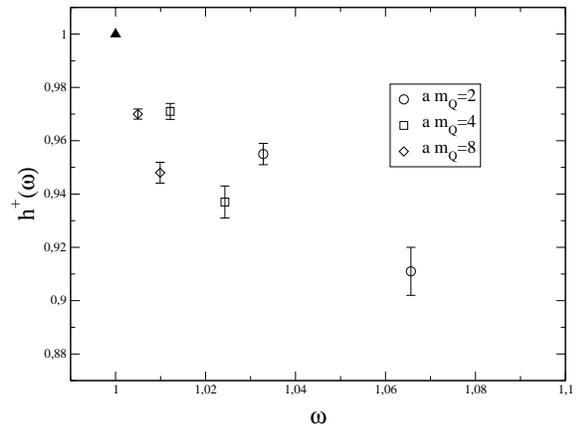}
\caption{Form factors as functions of $\omega$}
\label{fig:isgurwise}
\end{figure}

Now we can plot the form factors as function of $\omega=v\cdot v'$ , the
Isgur-Wise function~(Fig.~\ref{fig:isgurwise}). The errors are half of those
from using only one smearing. Each dataset extrapolates to the point $(1,1)$
as it has to for NRQCD. There seems to be a tendency of increasing slope with
the heavy quark mass which has to be explored further.

\section{Acknowledgement}
\label{sec:acknowledgement}

M.G. and J.H. acknowledge the financial support provided through the European
Community's Human Potential Programme under contract HPRN-CT-2000-00145,
Hadrons/Lattice QCD. This work was also supported by PPARC, NSF and DoE.

\end{document}